\documentclass[12pt]{iopart}

\usepackage[dvips]{graphicx}
\usepackage{rotating}
\newcommand{\be}{\begin{eqnarray}}
\newcommand{\ee}{\end{eqnarray}}

\def\BE{\begin{equation}}
\def\EE{\end{equation}}

\def\lsim{\mathrel{\raise.3ex\hbox{$<$\kern-.75em\lower1ex\hbox{$\sim$}}}}
\def\gsim{\mathrel{\raise.3ex\hbox{$>$\kern-.75em\lower1ex\hbox{$\sim$}}}}

\def\fun#1#2{\lower3.6pt\vbox{\baselineskip0pt\lineskip.9pt}}
\makeatletter

\def\vereq#1#2{\lower3pt\vbox{\baselineskip1.5pt \lineskip1.5pt
\ialign{$\m@th#1\hfill##\hfil$\crcr#2\crcr\sim\crcr}}}

\makeatother
\newcommand{\jpsi}{\ensuremath{J/\psi}}

\newcommand{\ups}{$\Upsilon$}

\newcommand{\pp}{p+p}           
\newcommand{\pbpb}{Pb+Pb}           
\newcommand{\PbPb}{Pb+Pb}           
\newcommand{\smallurl}[1]{{\small{\url{#1}}}}

\begin{document}

\title{Triggering on hard probes in heavy ion collisions with CMS}

\author{G Roland for the CMS Collaboration}

\address{Massachusetts Institute of Technology, Cambridge, MA 02139, USA}
\ead{rolandg@mit.edu}
\begin{abstract}
We present a study of the CMS trigger system in heavy-ion collisions.
Concentrating on two physics channels, dimuons from decays of quarkonia and single jets,
we evaluate a possible trigger strategy for \PbPb\ running that relies on event selection
solely in the High-Level Trigger (HLT).  The study is based on measurements of the
timing performance of the offline algorithms and event-size distributions using full simulations.
Using a trigger simulation chain,
we compare the physics reach for the jet and dimuon channels
using online selection in the HLT to minimum bias running. The results
demonstrate the crucial role the HLT will play for CMS heavy-ion physics.
\end{abstract}

\section{Trigger strategy in CMS \pbpb\ running \label{sec:hlt_intro}}

The key component in exploiting the CMS capabilities in heavy-ion collisions 
is the trigger system, which is crucial in accessing rare probes
such as high $E_T$ jets 
and photons, $Z^0$ bosons, $D$ and $B$ mesons, and
high-mass dileptons from the decay of quarkonia.  
The unique CMS trigger architecture only 
employs two trigger levels: The Level-1 trigger is implemented using custom electronics
and inspects events at the full bunch crossing rate. All further online selection is performed
in the High-Level Trigger (HLT) using a large cluster of commodity workstations (the ``filter farm'')
running ``offline'' reconstruction algorithms on fully assembled event information.
The trigger system was designed to deal with the unprecedented luminosities in LHC \pp\ running, yielding
an expected event rate of 40~MHz, with 25 superimposed \pp\ collisions for each event. Out of the 
40~MHz \pp\ event rate, a 100~kHz data stream will be selected by the Level-1 trigger for further processing
in the HLT. The HLT will reduce the 100~kHz input stream to 150~Hz of events written to permanent storage.

\begin{figure}[Hhtb]
\centering
\includegraphics[width=0.48\textwidth]{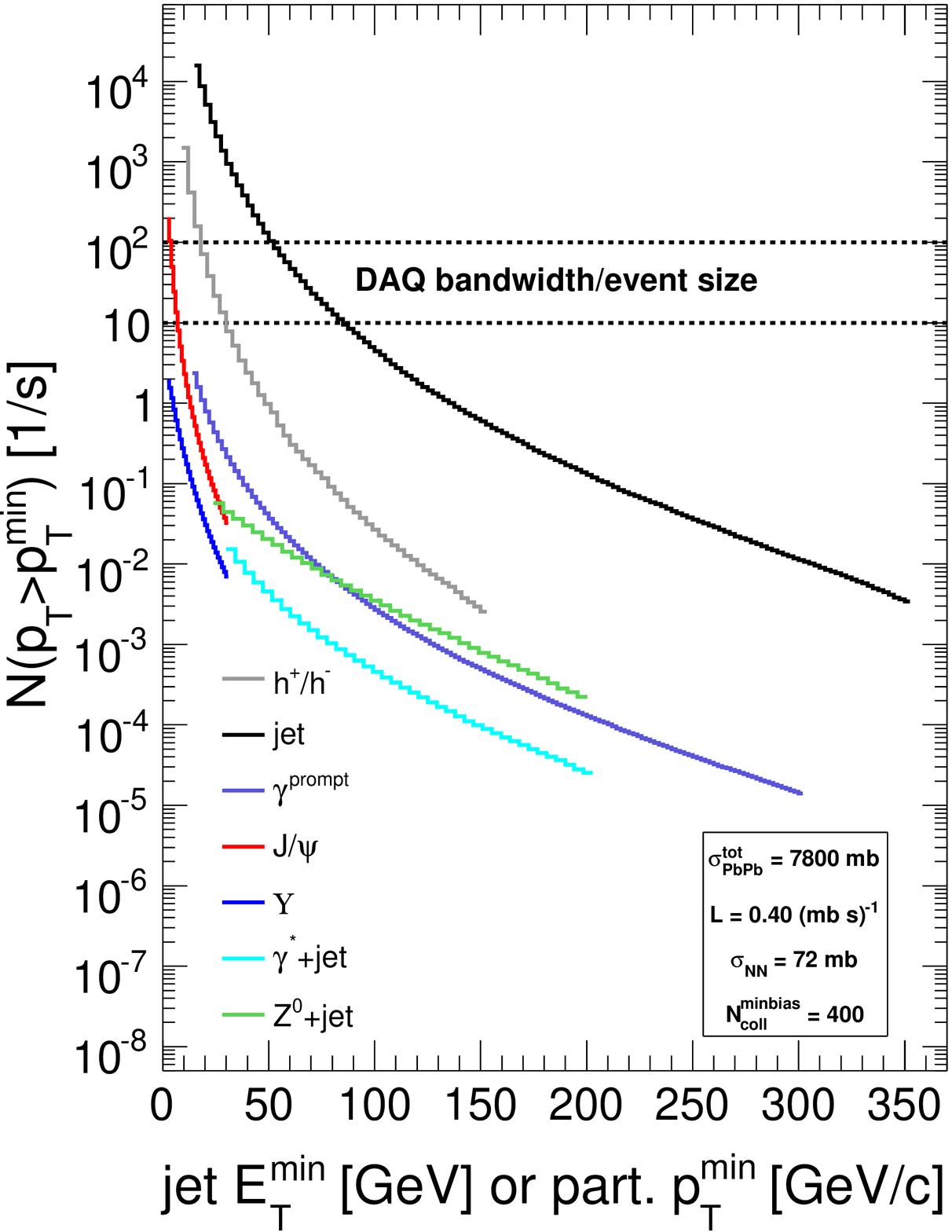}
\includegraphics[width=0.48\textwidth]{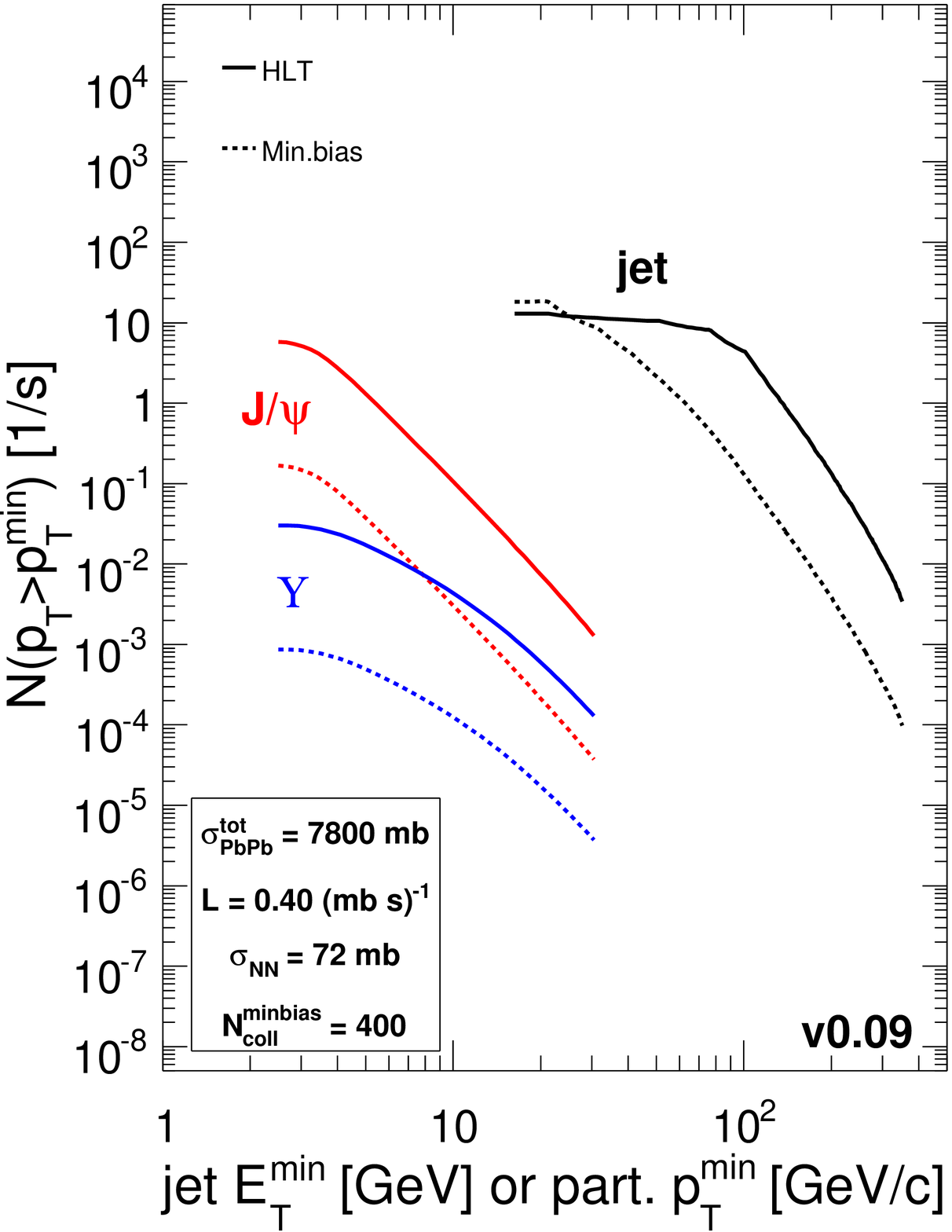}
\caption{\label{fig:prodrates} 
Left: Expected production rates in minimum bias \pbpb\ 
collisions at $\sqrt{s_{_{NN}}} = 5.5$~TeV, assuming design luminosity. Right: Expected rates to tape for jets, 
\jpsi\ and \ups\ channels for minimum bias (dashed lines) and HLT triggered data sets (solid lines).}
\end{figure}

The \pbpb\ design luminosity $L = 10^{27}$~cm$^{-2}$s$^{-1}$ at the beginning of a store is smaller than 
the \pp\ design luminosity by 7 orders of magnitude. 
The corresponding initial \pbpb\ collision rate is $\approx $8~kHz, 
and the average collision rate over the duration 
of a store will be $\approx $3~kHz. 
Therefore, even the maximal event rate for \pbpb\ collisions is much smaller than the 100~kHz input 
rate for the HLT in \pp\ collisions after the Level-1 selection. 
This allows a trigger strategy in \pbpb\ running that can 
be summarized as follows: Every \pbpb\ collision identified by the Level-1 trigger will be sent to the HLT filter farm. At the 
HLT, the full event information will be available for each event. All rejection
of \pbpb\ collisions will be based on the outcome of HLT trigger algorithms that are 
identical to the corresponding offline algorithms or optimized versions of the offline algorithms.
Therefore, algorithms like the offline jet finder will be run on each \pbpb\ event in the CMS interaction region,
optimizing the CMS physics reach.
This strategy relies on the fact that the HLT in its final configuration will provide sufficient input 
bandwidth to accept all collision events and sufficient computing power to run full offline algorithms
on all events.

The event size and computing time constraints were evaluated using full GEANT 4 based simulations
of \pbpb\ events. For events with a charged hadron multiplicity of $dN/d\eta \approx 3000$ for
central events, the event size was found to be approximately linear in 
the charged hadron multiplicity, ranging from 330~kByte/event for a $b=12$~fm sample to 8.5~MByte/event
for a $b=0$~fm sample. Averaging over impact parameter and adjusting for additional noise, backgrounds and 
diagnostic information, we obtain an event size of 2.5~MByte per minimum bias event for running at design 
luminosity.  
Including all uncertainties, we expect that the bandwidth of 225~MByte/sec will allow 
a rate of \pbpb\ events to mass storage between $10$ and $100$~Hz. 
A large part of this uncertainty will only be resolved once the first LHC data are taken, 
underscoring the need for a flexible high-level trigger scheme.

The HLT online computing farm is expected to consist of about $1500$ servers. 
The projected CPU budget per event, in units of todays 1.8~GHz Opteron CPUs on which our timing 
measurements were performed, will be $\approx$~1.5~s at the beginning of each store (8~kHz collision rate), 
and $\approx$~4~s averaged over the duration of the store (3~kHz collision rate). 

We measured the timing of three key algorithms in the CMS ORCA framework: the jet finding algorithm, the stand-alone muon 
finder using muon chamber information and the full muon finder including the silicon tracker information.
Averaging over the Glauber impact parameter 
distribution, the execution time of the modified iterative cone jet finding algorithm, including
background subtraction, is $\langle t \rangle = 250$~ms. 
More than 50\% of the execution time was spent in unpacking the calorimeter data. The stand-alone muon algorithm
has an average execution time of $\langle t \rangle = 80 \pm 20$~ms, using muon candidates from the 
Level-1 trigger. Both algorithms therefore fit comfortably into the CPU budget per event discussed above.

The full muon algorithm extends the tracks found in the muon system to the silicon tracker and provides a significant improvement 
in momentum resolution and background rejection. This is particularly important for low $p_T$ dimuons, which are expected to
take up the largest fraction of the output bandwidth to tape. This algorithm is called for less than 2\% of all events.
The current $L3$ execution time corresponds to about $10\pm 3$~s per minimum bias event.
Work on porting the the present offline algorithms to a new framework, CMSSW, and optimizing or significantly
modifying the present algorithms for use in the HLT event selection is ongoing. 

\begin{figure}[Hthb]
\begin{center}
\centerline{
\resizebox{75mm}{!}{\includegraphics{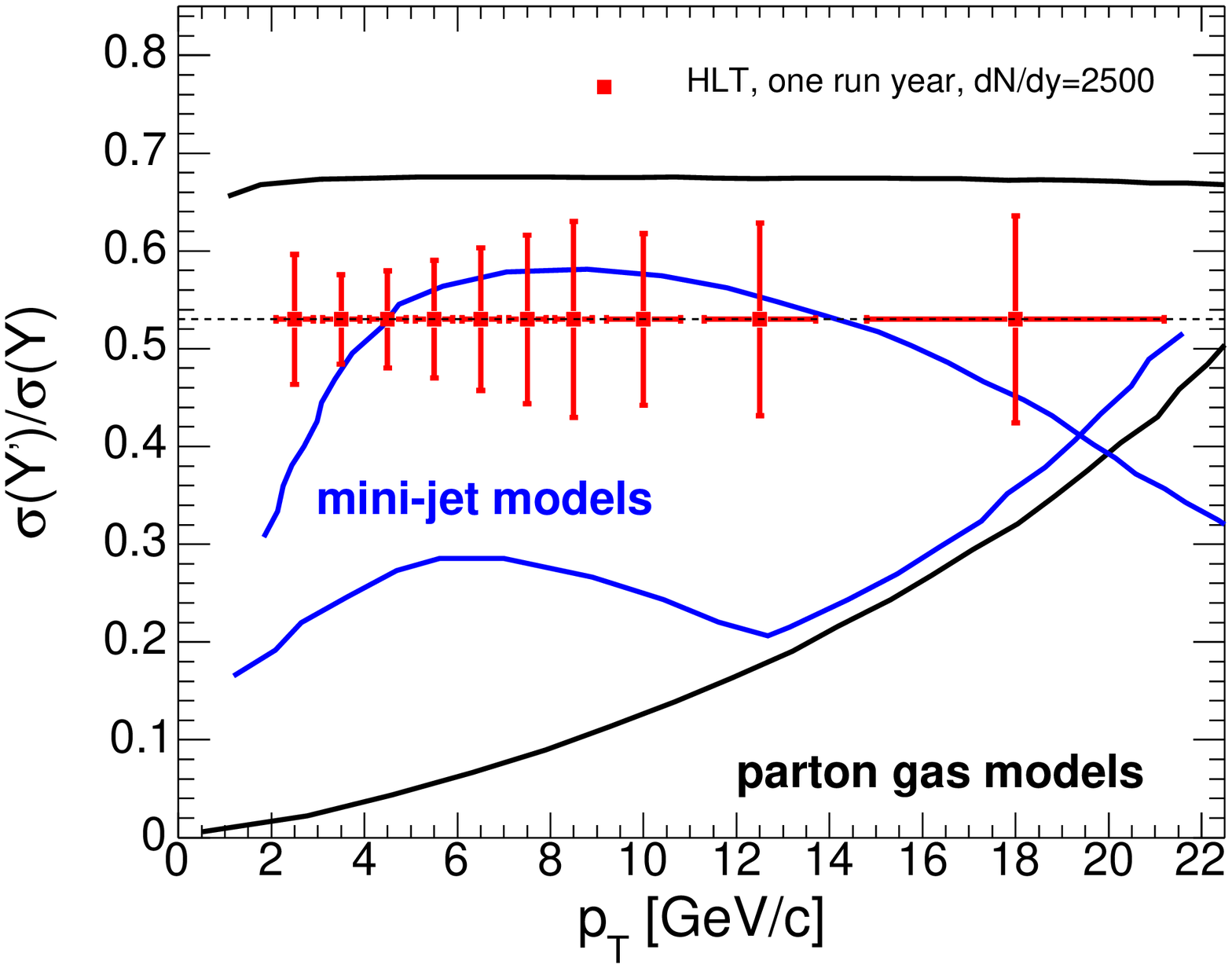}}
\resizebox{75mm}{!}{\includegraphics{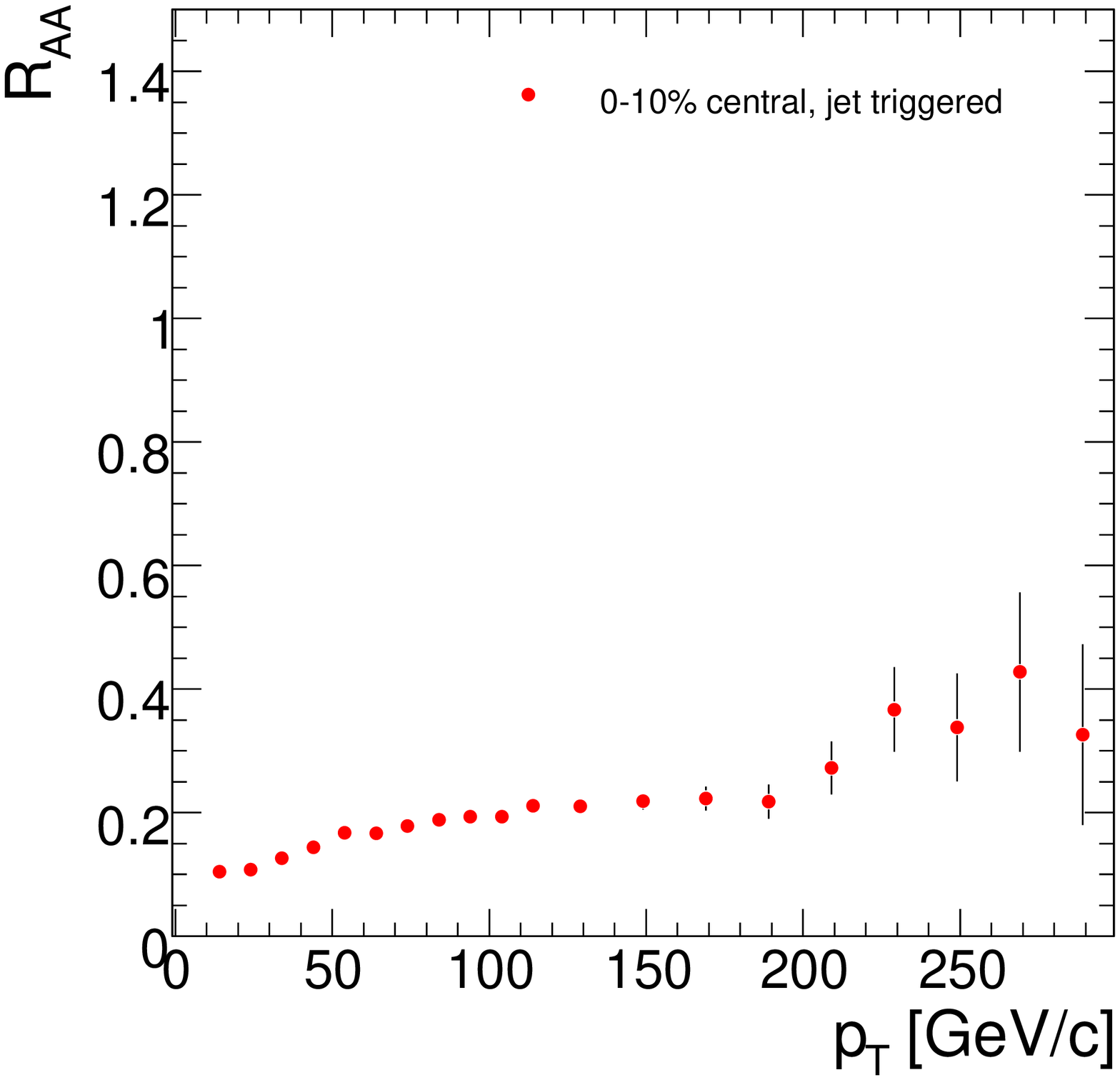}}}
\caption{\label{fig:raa}
Left: $\Upsilon$' over $\Upsilon$ ratio vs $p_T$. Statistics correspond
to $10^6$~s of data taking (one nominal year of LHC  heavy ion running). Shown in comparison to the statistical uncertainty
are calculations in different theoretical scenarios.
Right: The nuclear modification factor $R_{AA}$ as a function of $p_T$ for charged
particles, for minimum bias data (left panel) and for data triggered
on high-$E_T$ jets (right panel), for $10^6$~s of data taking.}
\end{center}
\end{figure}

\section{HLT simulation results  \label{sec:hlt_simulation}}
\label{SimChain}

Using results from event size and timing studies, a simulation chain was set up to translate the 
production cross-sections into rates to tape. The simulations used parametrizations of the acceptance,
efficiency and background rates of the offline jet finding and muon finding algorithms which are expected
to form the basis of the HLT algorithms. Details of the model calculations and reconstruction algorithms
can be found in \cite{hitdr}.

In Figure~\ref{fig:prodrates} (right) we show the rates of signal events to tape for 
minimum bias running (no event selection in HLT) in comparison to those for event selection using 
the HLT. The rates were calculated for design average luminosity, using a trigger table that 
devoted about 30\% of the output bandwidth to dimuon chanels and about 35\% to jet channels.

Using the HLT, a gain in statistics of more than an order of magnitude is achieved for jets at large $E_T$ and for dimuons. 
Correspondingly, the usable range in $E_T$ ($p_T$) for the jet and dimuon measurements is extended 
by more than than factor of $2$ and $3$, respectively.
Note that, for this comparison, the HLT rate for each process was only counted in the corresponding trigger stream.

Two key examples of the physics benefit of the HLT for quarkonium and jet related measurements are shown 
below. The left plot of Fig.~\ref{fig:raa} shows the ratio of ${\Upsilon}$' to $\Upsilon$ yields as a function
of transverse momentum. The projected statistical resolution is compared to four model calculations. This 
measurement, which relies on the added statistics provided by the HLT selection, allows a clear distinction of 
the different scenarios, and may therefore serve as a sensitive probe of the initial QCD medium.

In the right plot of Fig.~\ref{fig:raa} we show the nuclear modification factor $R_{AA}$ for 
events selected by an HLT trigger on high $E_T$ jets (right). Compared to a minimum bias data set,
the triggered sample extends the useful 
range in $p_T$ by more than a factor of 2.5, to more than 200~GeV/c. Predictions for  $R_{AA}$ in 
Pb+Pb collisions at the LHC have been made for several models of the parton energy loss in the 
QCD medium. The predictions differ most markedly in the high $p_T$ region, which can only be 
measured with high precision in the jet-triggered event sample.
In summary, the flexibility of the HLT system will allow us to 
allocate bandwidth to trigger channels differentially as a function of rapidity, $y$, and  $p_T$ 
of the trigger object and as a function of collision centrality, using full offline algorithms. 
This sophisticated triggering system will be critical in maximizing the overall physics reach 
of CMS in heavy ion running. 

%
%

\end{document}